\documentclass[showpacs,preprintnumbers,amsmath,amssymb,groupedaddress,
superscriptaddress]{revtex4}
\usepackage{graphicx}
\usepackage{dcolumn}
\usepackage{bm}

\begin{document}
\title{Contributions of different neutron pairs in different approaches for neutrinoless double beta decay}

\author{Alberto Escuderos }
\affiliation{Institute f\"{u}r Theoretische Physik der Universit\"{a}t
T\"{u}bingen, D-72076 T\"{u}bingen, Germany}
\author{Amand Faessler}
\email{amand.faessler@uni-tuebingen.de}
\affiliation{Institute f\"{u}r Theoretische Physik der Universit\"{a}t
T\"{u}bingen, D-72076 T\"{u}bingen, Germany}
\author{Vadim Rodin}
\affiliation{Institute f\"{u}r Theoretische Physik der Universit\"{a}t
T\"{u}bingen, D-72076 T\"{u}bingen, Germany}
\author{Fedor \v Simkovic}
\altaffiliation{On leave of absence from the BLTP, JINR, Dubna, Russia and
Department of Nuclear Physics, Comenius University, Mlynsk\'a dolina F1, SK--842 15
Bratislava, Slovakia}
\affiliation{Institute f\"{u}r Theoretische Physik der Universit\"{a}t
T\"{u}bingen, D-72076 T\"{u}bingen, Germany}

\date{\today}

\begin{abstract}
The methods used till now to calculate the neutrinoless double beta decay matrix elements are: the Quasiparticle Random Phase Approximation (QRPA), the Shell Model (SM), the angular momentum projected Hartee-Fock-Bogoliubov  approach (HFB) and the Interacting Boson Model (IBM). The different approaches are compared specifically concerning the the angular momenta and parities of the neutron pairs, which are changed into two protons by the $0\nu\beta\beta$-decay. The QRPA and SM involve about the same angular momentum and parity neutron pairs, while the HFB is restricted to $0^{+}, 2^{+}, 4^{+}, \dots$, and the IBM to $0^{+}$ and $2^{+}$ nucleon pairs.
\end{abstract}



\date{\today}

\nopagebreak[4]

\maketitle

\section{Introduction}

Neutrinoless double beta decay ($0\nu\beta\beta$ decay) allows to determine whether the neutrino is a Majorana or Dirac particle and gives also a value for the absolute scale of the neutrino masses and information on coupling constants and masses of physics beyond the Standard Model for Grand Unified Theories (GUT's), Supersymmetry (SUSY) and models with extra dimensions (see e.g. ~\cite{Doi85, Tomoda2, Simkovic98, Avig08}). The different $0\nu\beta\beta$-decay transition matrix elements are as important as the data on $0\nu\beta\beta$ decay to determine the neutrino mass and other parameters beyond the Standard Model. In the last three years the reliability of the transition matrix elements has been greatly improved (see e.g. ~\cite{Rod05,Ana,Suh05} ). The groups in T\"ubingen, Bratislava and Jyv\"{a}skyl\"{a}~\cite{Rod05,Ana,Suh05} are using mainly the Quasiparticle Random Phase approximation (QRPA), while Poves and collaborators~\cite{Poves1,Poves2,Poves3} use the Shell Model (SM), Tomoda, Faessler, Schmid and Gruemmer~\cite{Tomoda1} and Rath and coworkers~\cite{Rath1,Rath2,Rath3} use the angular momentum projected Hartree-Fock-Bogoliubov method (HFB), and Barea and Iachello~\cite{Iachello} use the Interacting Boson Model (IBM).

The QRPA ~\cite{Rod05, Suh05} has the advantage to allow to use a large single-particle basis. Thus, one is able to include to each orbit in the QRPA model space also the spin-orbit partner, which guarantees that the Ikeda sum rule ~\cite{Ikeda} is fulfilled, that is essential to describe correctly the Gamow-Teller strength. The SM~\cite{Poves1} is presently still restricted to a nuclear basis of four to five single-particle levels  for the description of double beta decay. Therefore, not all spin-orbit partners can be included and, as a result, the Ikeda sum rule is violated. On the other side, the shell model can in principle take into account all many-body configurations in a given single-particle basis. The excited states in the QRPA for spherical even-mass nuclei include `seniority' (the number of broken quasiparticles) states two, six, ten, \dots and for the ground state correlations `seniority' zero, four, eight, \dots configurations. The SM takes for the ground state seniority zero, four, six, eight, \dots, and for the excited states seniority two, four, six, \dots into account. Since the SM uses the closure 
relation for calculating the $0\nu\beta\beta$ matrix elements, one needs only the configurations for the initial and final even-even nuclei in the $0^{+}$ ground states. In contrast to the QRPA, the SM includes also states of the seniority six, but on the other side the SM violates the Ikeda sum rule due to the small single-particle basis. The effects of the seniority six configurations and of the violation of the Ikeda sum rule are also studied in this work.

The difference of the HFB and the IBM approach to the QRPA and the SM treatment is for the later two the higher flexibility to include more angular momenta and parities for the two neutron pairs, which are changed into two protons in the $0\nu\beta\beta$ transitions. The HFB approach with angular momentum projection after variation allows only neutron pairs with $J^\pi=0^{+}, 2^{+}, 4^{+}, \dots$ or in general $ (-)^{J} = 1$,  positive parity states for real coefficient and no parity mixing  in the Quasiparticle transformation~\cite{Schmid1, Schmid2} (see equation (\ref{pair1})). The IBM is restricted to neutron ``S'' and ``D'' pairs, i. e.  two neutron states with angular momenta $0^{+}$ and $2^{+}$, which is a restriction not found in the more flexible QRPA and SM approaches as shown in this work.

\section{The different Many Body Approaches and the transformed Neutron Pairs}

Here a short characterization to the different many body approaches: QRPA~\cite{Rod05, Ana, Suh05}, 
SM ~\cite{Poves1, Poves2}, 
HFB ~\cite{Schmid1, Schmid2, Rath1, Rath2} 
and the IBM ~\cite{Iachello} 
is given.

In the QRPA one starts from the transformation to Bogoliubov quasiparticles :

\begin{equation}
a_{i}^{\dagger} = u_{i} c_{i}^{\dagger} - v_{i} c_{\bar{i}}.
\end{equation}

The creation $c_{i}^{\dagger}$ and annihilation operators of time reversed single-particle states $c_{\bar{i}}$ are usually defined with respect to oscillator wave functions~\cite{Rod05}. The single-particle energies are calculated with a Woods Saxon potential~\cite{Rod05}. The single-particle basis can include up to 23 nucleon levels (all single-particle states without a core up to the $i_{13/2}$ level) for the protons and also for the neutrons. But the QRPA results for the $0\nu\beta\beta$ matrix elements turn out to be stable as a function of the basis size already for smaller basis sets (from 6 or 7 levels and larger, respectively).

The excited states $|m\rangle$ with angular momentum $J$ in the intermediate odd-odd mass nucleus are created from the correlated initial and final $0^{+}$ ground states by a proton-neutron phonon
creation operator:
\begin{equation}
|m\rangle=Q_{m}^{\dagger} |0^{+}\rangle; \ \ \hat{H} Q_{m}^{\dagger} |0^{+}\rangle = E_{m} Q_{m}^{\dagger} |0^{+}\rangle.
\label{H}
\end{equation}
\begin{equation}
Q_{m}^{\dagger} = \sum_{\alpha} [ X_{\alpha}^{m} A_{\alpha}^{\dagger} - Y_{\alpha}^{m} A_{\alpha}],
\label{Q}
\end{equation}
which is defined as a linear superposition of creation operators of proton-neutron quasiparticle pairs:
\begin{equation}
A_{\alpha}^{\dagger} = [ a_{i}^{\dagger} a_{k}^{\dagger}]_{J M},
\label{BCS0}
\end{equation}
\begin{equation}
[A_{\alpha}, A_{\beta}^{\dagger}] = \delta_{\alpha , \beta} + \hat{x},
\label{X}
\end{equation}

The ``scattering" terms $\hat{x}$ in (\ref{X}) are put to zero in the QRPA and are included approximately in Renormalized QRPA (RQRPA), which for the first time was employed to neutrinoless double beta decay in the PhD thesis of Schwieger in T\"ubingen~\cite{Schwieger}. The RQRPA includes the Pauli principle for the Fermion pairs and reduces the ground state correlations. The many body Hamilton equation (\ref{H}) yields the algebraic QRPA equation for the the coefficients $X_{\alpha}^{m}$  and $Y_{\alpha}^{m}$, and the excitations energies $E_{m}$. The approach with $\hat{x} \equiv 0 $ is the so called Quasi Boson Approximation (QBA).

The inverse $0\nu\beta\beta$ lifetime for the light Majorana neutrino exchange mechanism is given as the product of three factors,
\begin{equation}
\label{T1/2}
\left(T^{0\nu}_{1/2}\right)^{-1}=G^{0\nu}\,\left|M^{0\nu}\right|^2\,m_{\beta\beta}^2\ ,
\end{equation}
where $G^{0\nu}$ is a calculable phase space factor, $M^{0\nu}$ is the $0\nu\beta\beta$ nuclear
matrix element, and $m_{\beta\beta}$ is the (nucleus-independent)
``effective Majorana neutrino mass'' which, in standard notation \cite{PDGr}, reads

\begin{equation}
m_{\beta\beta}=\left| \sum_{l=1}^3 m_l  U_{el}^2 \right|\ ,
\label{mn}
\end{equation}


with $m_l$ and $U_{el}$ being the neutrino masses 
and the $\nu_e$ mixing matrix elements, respectively.

The expressions for the matrix elements $M^{0\nu}$ and the corresponding $0\nu\beta\beta$ transition operators are given, e.g., in Ref.~\cite{Rod05}:
\begin{equation}
M^{(0\nu)} =  M_{GT}^{0\nu} - ( \frac{g_{V}}{g_{A}})^2 M_F^{0\nu} - M_T ^{0\nu}
\label{Mnu}
\end{equation}
\begin{equation}
M_K^{0\nu} = \langle 0^{+}_f| H_K(r_{12}, E^m_{J^{\pi}})|0^+
\rangle, \ \ \  H_K(r_{12}, E^m_{J^{\pi}}) = \frac{2R}{\pi g_A^2}\int_0^{\infty}f_K(qr_{12}) \frac{h_K(q^2) q dq}{q + E^m_{J^{\pi}} - (E_i -  E_f)/2}.
\label{MK}
\end{equation}
Here, ``K'' labels GT (Gamow Teller), F (Fermi) and T (Tensor) contribution. The functions $f_K(qr_{12})$ are spherical Bessel functions $j_0(qr_{12}) $ for F and GT and $j_2(qr_{12})$ for T. The function $h_K(q^2)$ is defined in appendix A of Simkovic et al.~\cite{Ana}, and $E^m_{J^{\pi}}$ are the energies of states in the intermediate nucleus. The nuclear radius is $R=r_0A^{1/3}$, and  the parameter $r_0$ is chosen here to be $r_0=1.2$ fm. Thus, the matrix elements $M^{0\nu}$ are increased by about 10 \%  as compared with our earlier calculations, where we chose $r_0=1.1$ fm.

\begin{figure}[htb]
\includegraphics[scale=0.3]{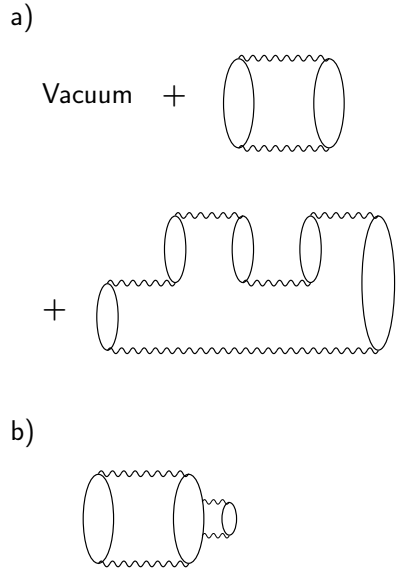}~~
\caption{The ring diagrams of type a) are with the ``vacuum" included in the ground state correlations in the QRPA. This means that the seniority, or in deformed nuclei the quasiparticle number, zero, four, eight, twelve, \dots  excitations are included in both the QRPA and the SM. But, e.g., diagram  b) is not included in the QRPA but is considered in the SM, where seniority 4, 6, 8, 10, 12, \dots are included in the ground state correlations~\cite{Poves1, Poves2, Poves3}. However, contributions of the seniority six and ten admixtures to the neutrinoless double beta decay probability are suppressed according to the philosophy of the Random Phase Approach, that the ring diagrams are the leading ones (see also figure 4).}
\label{ring}
\end{figure}

\begin{figure}[htb]
\includegraphics[scale=0.6, angle=-90
]{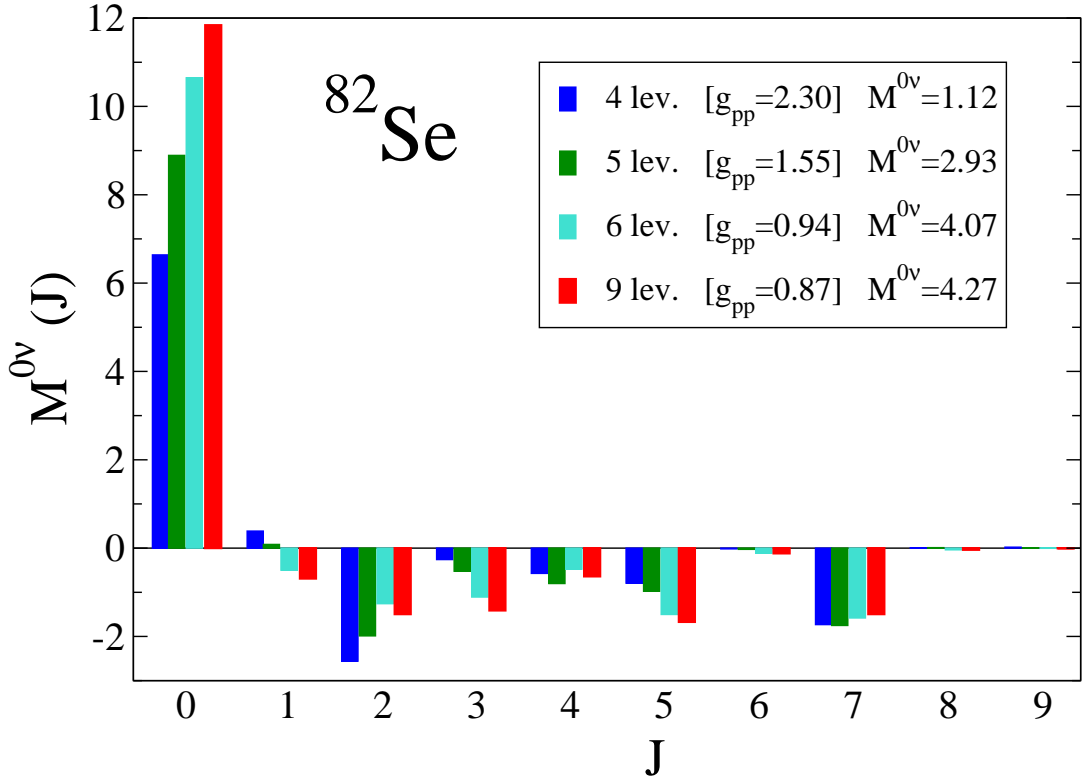}~~~~
\caption{(Color online) Contributions of the transforming neutron pair with
different angular momenta $J^\pi$ to the total $M^{0\nu}$ calculated within the QRPA and different basis sizes for the $0\nu\beta\beta$ decay $^{82}$Se$\to^{82}$Kr. The left bar is calculated with the same basis of four levels, $1p_{3/2}, 0f_{5/2}, 1p_{1/2}$ and $0g_{9/2}$, used in the shell model calculations. The Ikeda Sum Rule (ISR) is exhausted by 50\%. The second bar from the left includes in addition one, the $1f_{7/2}$ level, of the two missing spin-orbit partners given for the $^{82}$Se nucleus in ref. ~\cite{Poves4} for the shell model.  The ISR is exhausted by 66\%.  The third bar from the left  includes both missing  spin-orbit partners $0f_{7/2}$ and $0g_{7/2}$ amounting in total to 6 single-particle levels. The ISR is fulfilled by 100\%.  This leads to the increase in the neutrinoless matrix element from 1.12 to 4.07. The right bar represents the QRPA result with 9 single-particle levels $ (1f_{7/2}, 2p_{3/2}, 1f_{5/2}, 2p_{1/2}, 1g_{9/2}, 2d_{5/2}, 3s_{1/2}, 2d_{3/2}, 1g_{7/2}.)$. The matrix element gets only slightly increased to 4.27. The spin-orbit partners are essential to fulfill  the Ikeda Sum Rule (ISR).
In all four QRPA calculations the QRPA ``renormalisation'' factor $g_{pp}$  (given in the figure) of the particle-particle strength of the Bonn CD nucleon-nucleon interaction is adjusted to reproduce the experimental $2\nu\beta\beta$ decay rates.}
\label{QRPA-SM1}
\end{figure}

\begin{figure}[htb]
\includegraphics[scale=0.6
]{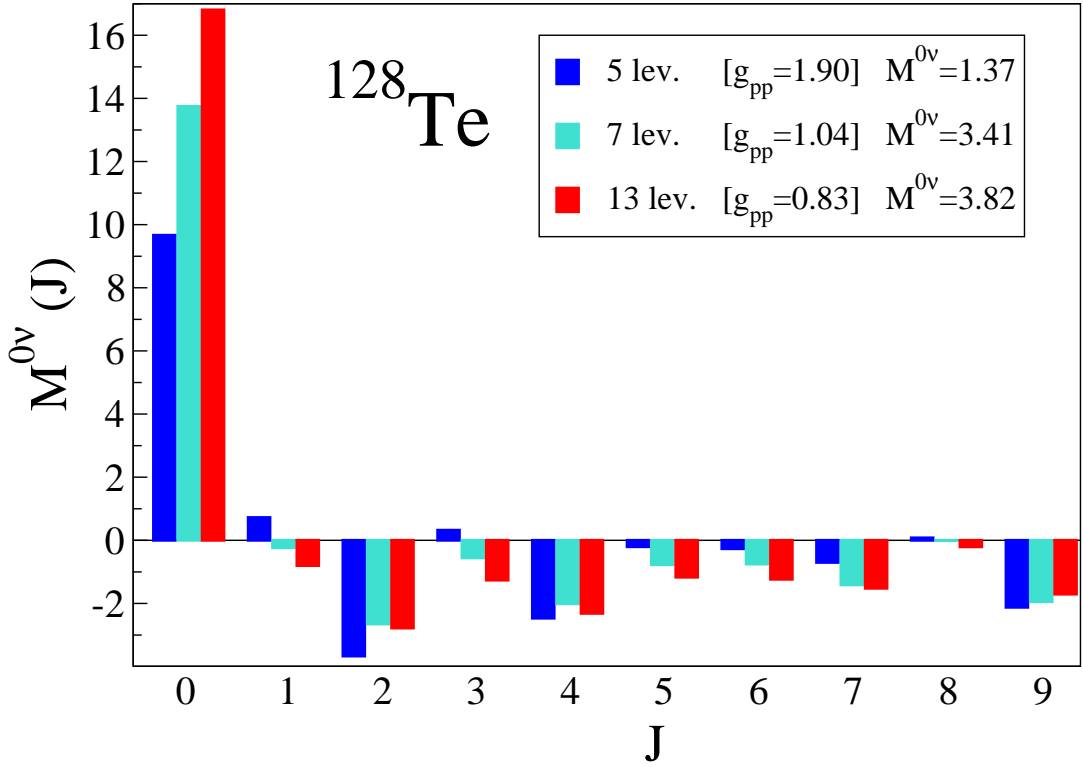}~~~~
\caption{(Color online) Contributions of different angular momenta associated with the transforming neutron pair for the QRPA with different basis sizes and with
different angular momenta $J^\pi$ for the $0\nu\beta\beta$ decay $^{128}$Te$\to ^{128}$Xe. The left bar is calculated with the same five single-particle levels, $0g_{7/2}, 1d_{5/2}, 2s_{1/2}, 1d_{3/2}$ and $0h_{11/2} $, used in the shell model calculations. The middle bar includes in addition the missing spin-orbit partners $0g_{9/2}$ and $0h_{9/2}$, with 7 levels in total. This increases the neutrinoless matrix element from $1.37$ to $ 3.41$. The right bar is the QRPA result obtained with 13 single-particle levels including all states from the $ N = 3 (p,f)$ and the $ N = 4 (s,d,g)$ shells and the four additional levels $0h_{11/2}, 0h_{9/2}, 1f_{7/2} $ and $ 1f_{5/2}$ from the $ N = 5 $ shell. The matrix element is only increased slightly to $ 3.82 $. The spin-orbit partners are essential to fulfill the Ikeda sum rule. The further increase of the basis from 7 to 9 levels produces only a small change from $3.41$ to $ 3.82 $ for the $0\nu\beta\beta$ transition matrix element. In all three calculations the QRPA ``renormalisation'' factor $g_{pp}$  (given in the figure) of the particle-particle strength of the Bonn CD nucleon-nucleon interaction is adjusted to reproduce the experimental $2\nu\beta\beta$ decay rates.
}
\label{QRPA-SM2}
\end{figure}

The SM approach has been applied by the Strasbourg-Madrid group ~\cite{Poves2} to  neutrinoless double beta decay ~\cite{Poves1}  using the 
closure relation with an averaged energy denominator. In this way one does not need to calculate the states in the odd-odd intermediate nuclei. The quality of the results depends then on the description of the $0^{+}$ ground states in the initial and final nuclei of the double beta decay system, e.g. $^{76}$Ge $\to ^{76}$Se. The $0\nu\beta\beta$ transition matrix element (\ref{Mnu}) simplifies as shown in Ref.~\cite{Poves3} equations (5) to (11). Since the number of many body configurations is increasing drastically with the single-particle basis, one is forced to restrict for mass numbers $A = 76$ and $A = 82$ in the SM to the single-particle basis $1p_{3/2}, 0f_{5/2}, 1p_{1/2}$ and $0g_{9/2}$. In ref.~\cite{Poves4} the $^{82}$Se nucleus is calculated in the SM for five basis single-particle levels including also $0f_{7/2}$. For the mass region around $A = 130$ the SM basis is restricted to $0g_{7/2}, 1d_{3/2}, 1d_{5/2}, 2s_{1/2}$  and  $0h_{11/2}$ levels. The problem with these small basis sets is that the spin-orbit partners $0f_{7/2}$ and $0g_{7/2}$ have to be omitted~\cite{Poves3}. The SM results then automatically violate the Ikeda Sum Rule (ISR)~\cite{Ikeda}, while the QRPA satisfies it exactly. The Ikeda sum rule is:
\begin{equation}
S_- - S_+ = 3(N - Z),
\label{Ikeda}
\end{equation}
\begin{equation}
S_- = \sum_{\mu} \langle 0^{+}_i|[ \sum_{k}^{A} (-)^{\mu}\sigma_{-\mu}(k) t_{+}(k)] [ \sum_{l}^{A} \sigma_{\mu}(l) t_{-}(l)] |0^{+}_i \rangle,
\label{S}
\end{equation}
For $S_{+}$ the subscripts at the isospin rising and lowering operators are exchanged.

\begin{figure}[htb]
\includegraphics[scale=0.65
]{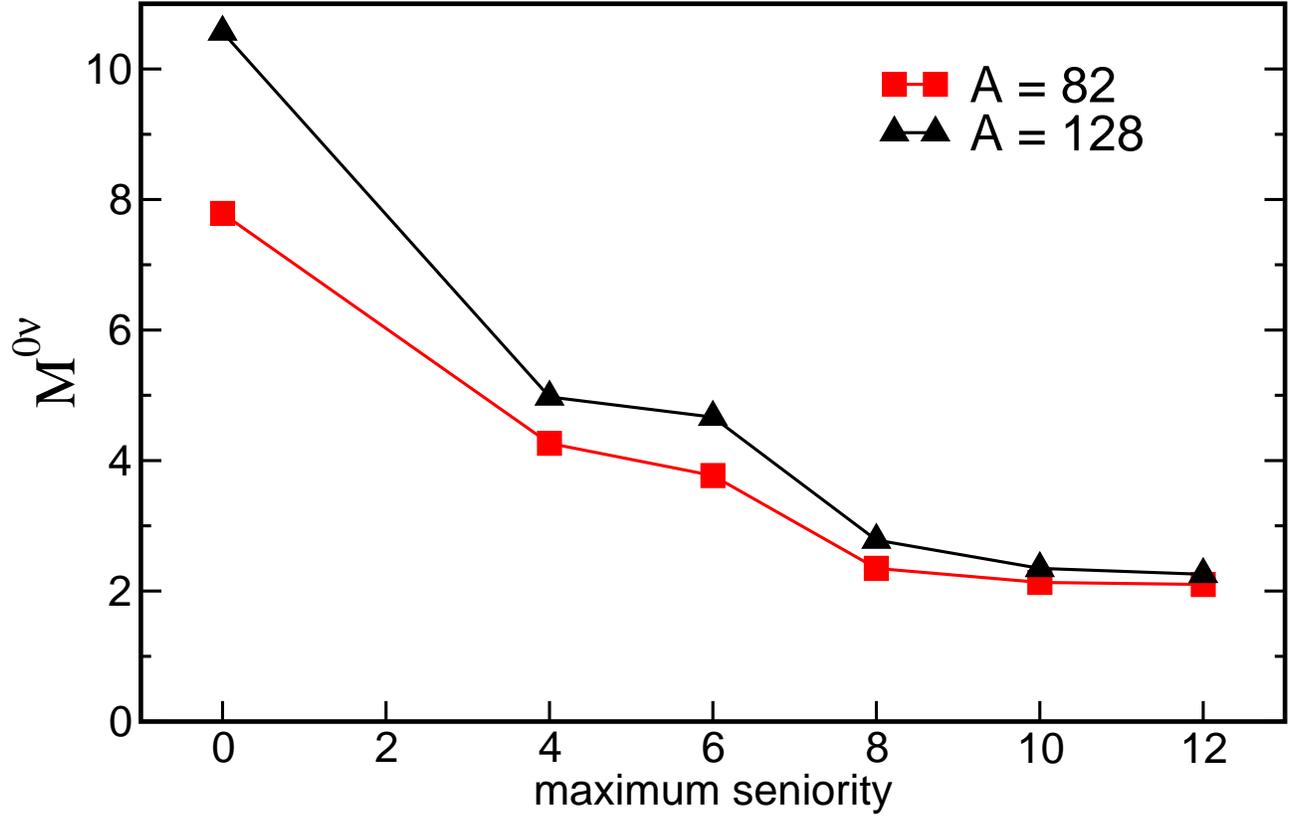}
\caption{(Color online) The $0\nu\beta\beta$ matrix element as a function of maximum seniority included in the SM wave functions for $^{82}$Se-$^{82}$Kr and of $^{128}$Te-$^{128}$Xe. The contributions of seniority 6 and 10 states are appreciably smaller in agreement with the philosophy of the Random Phase Approximation. The data in the figure are taken from Caurier et al.~\cite{Poves1}.}
\label{Poves-Fig}
\end{figure}

The SM is in principle a better approach, if the single-particle basis is large enough. The ground state correlation in the QRPA are limited to the ring diagrams (see Fig.~\ref{ring}), and thus to `seniority' (the number of broken quasiparticle) $ 0, 4, 8, 12, \dots $ configurations, while the shell model can include seniority  $0, 4, 6, 8, 10, 12, \dots $ states. This has been discussed in publications of the Strasbourg-Madrid group ~\cite{Poves1, Poves3}. But the results~\cite{Poves1, Poves3} show also that the philosophy of the RPA is approximately correct, since the ring diagrams give the most important ground state correlations. The contributions of seniority 6 and 10 to $M^{0\nu}$ are suppressed compared to the others (see figure ~\ref{Poves-Fig}).

Figures~\ref{QRPA-SM1} and~\ref{QRPA-SM2} show the QRPA contributions of different angular momenta of the neutron pairs, which are changed in  proton pairs with the same angular momenta. In figure ~\ref{QRPA-SM1} the left bar is the result for $^{82}$Se obtained with the single-particle basis $1p_{3/2}, 0f_{5/2}, 1p_{1/2}$ and $0g_{9/2}$ used in the SM.  The ISR is exhausted by 50\%. The second bar from the left represents the result with addition of the $1f_{7/2}$ level. The ISR is exhausted by 66\%. The third bar from the left shows the result obtained by inclusion of both  spin-orbit partners $0f_{7/2}$ and $0g_{9/2}$ missing in the four level basis. The ISR is 100\% fulfilled. For the right bar the basis is increased to 9 single-particle levels for neutrons and protons ($ 0f_{7/2}, 1p_{3/2}, 0f_{5/2}, 1p_{1/2}, 0g_{9/2}, 1d_{5/2}, 2s_{1/2}, 1d_{3/2}, 0g_{7/2}$).

For $^{128}$Te in figure ~\ref{QRPA-SM2} the left bar is calculated with the same five single-particle levels, $0g_{7/2}, 1d_{5/2}, 2s_{1/2}, 1d_{3/2}$ and $0h_{11/2}$,  used in the shell model calculations. The middle bar represents the results by inclusion in addition the missing spin-orbit partners $0g_{9/2}$ and $0h_{9/2}$, with 7 levels in total. The ISR is 100\% fulfilled. This strongly increases the $0\nu\beta\beta$ matrix element from $1.37$ to $ 3.41$. The right bar is the QRPA result with 13 single-particle levels including all the states from the $ N = 3 \  (p,f)$ and the $ N = 4 \ (s,d,g)$ shells and the four additional levels $0h_{11/2}, 0h_{9/2}, 1f_{7/2} $ and $ 1f_{5/2}$ from the $ N = 5 $ shell. The matrix element is increased only slightly to $ 3.82 $. The spin-orbit partners are essential to fulfill  the Ikeda Sum Rule (ISR). The further increase of the basis from 6 to 9 levels produces a small change only from $3.41$ to $ 3.82 $ for the $0\nu\beta\beta$ transition matrix element. For all results in figures 2 and 3  the $2\nu\beta\beta$ the QRPA ``renormalisation'' factor $g_{pp}$ of the particle-particle strength of the Bonn CD nucleon-nucleon interaction is adjusted to reproduce the experimental $2\nu\beta\beta$ decay rates. The radius parameter is chosen to be $r_0=1.2$ fm.

Figure ~\ref{Poves-Fig} shows for $^{82}$Se and for $^{128}$Te the influence of different seniority states in the SM ~\cite{Poves1}. A seniority zero pair corresponds to a zero angular momentum neutron pair, which can be changed into a zero seniority proton pair. It yields a large positive contribution to the $0\nu\beta\beta$ matrix element. The higher seniority states reduce the matrix elements  as seen also in figures 2 and 3 for the QRPA. One remarks, that seniority 6 and seniority 10 states, which are not included in the QRPA, give only a very small or even zero contribution to the $0\nu\beta\beta$ matrix elements in agreement with the RPA philosophy.

The QRPA is restricted to the ring diagrams for the ground state correlations, while the SM takes into account all possible-many body states in a given single-particle basis. The drawback of the SM is the limited basis size. For the light double beta decay systems like $^{76}Ge$ and $^{82}Se $ the single-particle basis is restricted to the four levels ~\cite{Poves1, Poves4, Poves2, Poves3} \  $1p_{3/2}, 0f_{5/2}, 1p_{1/2}, 0g_{9/2}$. (In ref.~\cite{Poves4} the $1f_{7/2}$ level is also included for $^{82}$Se.)  Since the spin-orbit partners $0f_{7/2}, 0g_{7/2} $ are missing, the Ikeda Sum Rule (ISR) is violated and one obtains only  50\% (66\% for five levels) of the the ISR (see table ~\ref{tab1}). Including the spin-orbit partners into the basis $ 0f_{7/2}, 1p_{3/2}, 0f_{5/2}, 1p_{1/2}, 0g_{9/2}, 0g_{7/2}$ the ISR is fulfilled. It has been argued, that including the spin-orbit partner $0f_{7/2}, 0g_{7/2} $ and moving the single-particle energy levels of these states to infinity also fulfills the ISR. This is correct, if one sums the states in the ISR to infinity. We have done a calculation with 6 levels moving $0f_{7/2}, 0g_{7/2} $ to 500 MeV. The ISR is fulfilled as expected, if one sums to energies larger than 500 MeV. However, the contributions to the ISR from low-lying states below about 10 MeV are within 1 \% the same as for the four level basis, and if one sums now the contributions to the ISR only from the physical states up to about 10 MeV, one obtains again only about 50 percent of the ISR. 
The nuclear structure properties of a six level basis with shifts of the two spin-orbit partners are markedly different from the calculation with the six levels with the correct single-particle energies.

Table I shows the contributions from `seniority' (the number of broken quasiparticles) zero and from `seniority' different from zero to the total matrix element $M^{0\nu}$ for different sizes of the single-particle basis in the QRPA calculations for $^{82}$Se, $^{128}$Te, $^{130}$Te and for the large basis in $^{76}$Ge, and for the small basis with 4 or 5 levels in the SM calculations~\cite{Poves1, Poves4, Poves2, Poves3} for the two nuclei $^{82}$Se, $^{128}$Te. In all the QRPA calculations the ``renormalisation factor'' $g_{pp}$ of the particle-particle (pp) force multiplying the pp-nucleon-nucleon matrix elements of the  G matrix of the Bonn CD interaction is adjusted to reproduce the experimental $ 2\nu\beta\beta $ decay rates. In Ref.~\cite{Lisi} the factor $g_{pp}$ has been adjusted together with the axial-vector charge $g_A$ simultaneously to the $2\nu\beta\beta$ decay, to the single $ \beta^-$ decay in the second leg and to electron capture in the first leg for three double beta decay systems. The results for the $0\nu\beta\beta$-decay matrix elements change with this overconstrained fit less than the general theoretical uncertainties (see figure ~\ref{total}). Only the Gamow-Teller transitions via the $1^+$ intermediate states depend sensitively on $g_{pp}$, since the $ 1^{+}$ states in the intermediate nucleus are close to a ``phase transition" to "static Gamow-Teller deformations". As a function of the other particle-particle multipole parts of the force the results change only very weakly. 
Therefore, possible uncertainties of $g_{pp}$, which affect a small contribution of the $1^+$ multipole, do not affect appreciably the total $0\nu\beta\beta$ matrix element.

\begin{table}[h]
\caption{Contribution from seniority zero and seniorities different from zero to the total value of the $0\nu\beta\beta$ transition matrix elements obtained for different nuclei in the QRPA and the SM with different basis sizes. The last column gives the exhaustion of the Ikeda Sum Rule (ISR) in percents (see eqs. (\ref{Ikeda}), (\ref{S})). The basis sets used are: 4 levels: $1p_{3/2}, 0f_{5/2}, 1p_{1/2}, 0g_{9/2}$ ; 5 levels: $ 0g_{7/2}, 1d_{5/2}, 2s_{1/2}, 1d_{3/2}, 0h_{11/2} $;
6 levels: $ 0f_{7/2}, 1p_{3/2}, 0f_{5/2}, 1p_{1/2}, 0g_{9/2}, 0g_{7/2}$; 7 levels: $ 0g_{7/2}, 1d_{5/2}, 2s_{1/2}, 1d_{3/2}, 0h_{11/2}, 0g_{9/2}, 0h_{9/2}$;
9 levels: $ 0f_{7/2}, 1p_{3/2}, 0f_{5/2}, 1p_{1/2}, 0g_{9/2}, 1d_{5/2}, 2s_{1/2}, 1d_{3/2}, 0g_{7/2}$ ;13 levels: All levels of $ N = 3 \  (p,f)$ and the $ N = 4 \ (s,d,g)$ shells and  $0h_{11/2}, 0h_{9/2}, 1f_{7/2} $ and $ 1f_{5/2}$.  Four levels (for: $ ^{76}$Ge, $^{82}$Se)  and five levels (for $^{100}$Mo, $^{128}$Te, $^{130}$Te) correspond to the SM basis sets ~\cite{Poves1, Poves4, Poves2, Poves3}. Since the level four and five basis sets do not contain all spin orbit partners, the ISR is strongly violated and only exhausted by 50 and 60 percents for these sets, respectively.}
\vspace{0.5cm}

\begin{tabular}{|r|l|c|c|c|c|c|}
	\hline	
Nucleus  &  Basis Size & Method   &  seniority: $ s = 0$ &         seniority: $ s \neq  0$ &  \ \   total\ \ & Ikeda SR percents \\
\hline
 $^{76}$Ge  & \ \ 9 & \ \  QRPA & \ \ 12.5 \ \ & -7.5 & 5.0 &  100 \\
 $^{82}$Se  &\ \  4 & \ \  QRPA & \ \ 6.6\ \        & -5.5 & 1.1 & 50 \\
 $^{82}$Se  & \ \ 4 & \ \  SM &    \ \ 7.8 \ \      & -5.8  & 2.0 &   \\
 $^{82}$Se  & \ \ 5 & \ \  QRPA & \ \ 8.9 \ \    & -6.0  & 2.9 &  66  \\
 $^{82}$Se  & \ \ 5 & \ \ SM &     \ \ $+f_{7/2}$ \ \ &   & 2.5 &      \\
 $^{82}$Se  & \ \ 6 & \ \  QRPA & \ \ 10.7\ \  & -6.6 & 4.1 & 100  \\
 $^{82}$Se  & \ \ 9 & \ \ QRPA &  \ \ 11.8  \ \ & -7.5 & 4.3& 100  \\
 $^{100}$Mo & \ \ 13 & \ \ QRPA & \ \ 16.3 \ \ & -12.6 & 3.7 & 100 \\
 $^{128}$Te  &\ \  5 & \ \  QRPA & \ \ 9.7\ \        & -8.3 & 1.4 & 60 \\
 $^{128}$Te  &\ \  5 & \ \  SM  & \ \ 10.6\ \        & -8.4 & 2.2 &   \\
 $^{128}$Te  & \ \ 7 & \ \  QRPA & \ \ 13.7\ \  & -10.3 & 3.4 & 100  \\
 $^{128}$Te  & \ \ 13 & \ \ QRPA &  \ \ 16.8  \ \ & -13.0 & 3.8& 100  \\
 $^{130}$Te  &\ \  5 & \ \  QRPA & \ \ 8.8\ \        & -7.5 & 1.3 & 60 \\
 $^{130}$Te  & \ \ 7 & \ \  QRPA & \ \ 12.2\ \  & -9.0 & 3.2 & 100  \\
 $^{130}$Te  & \ \ 13 & \ \ QRPA &  \ \ 14.9  \ \ & -11.1 & 3.8 & 100  \\

\hline
\end{tabular}
\label{tab1}
\end{table}

\begin{table}[h]
\caption{Contributions (in \%) of different `seniorities'  (the number of broken quasiparticles)  to the normalization of the QRPA ground states for different nuclei. The 9 and 13 level single-particle basis sets are defined in the caption of table I. The renormalisation factor  $g_{pp}$ of the Bonn CD NN force is adjusted to reproduce the experimental $2\nu\beta\beta$  half lives. A larger basis and also a stronger NN force produces larger admixtures of higher `seniorities'. }
\vspace{0.5cm}

\begin{tabular}{|c|c|c|c|c|c|c|}
\hline

nucleus  & basis & \  $ s=0$  \  & \ $s=4$ \   & \  $s=8$\  & \  $s=12$ \  & \ $s=16$ \  \\
    \hline

$^{76}$Ge  & 9 &    55.0 &  32.9 &  9.8  &  2.0  &  0.3 \\
$^{76}$Se  & 9 &    58.7 &  31.3 &  8.3  &  1.5  &  0.2 \\
$^{82}$Se  & 9 &    56.4 &  32.3 &  9.3  &  1.8  &  0.3 \\
$^{82}$Kr  & 9 &    53.5 &  33.5 &  10.5 &  2.2  &  0.3 \\
$^{128}$Te & 13 &   51.5 &  34.2 &  11.3 &  2.5  &  0.4 \\
$^{128}$Xe & 13 &   39.9 &  36.7 &  16.9 &  5.2  &  1.2 \\

\hline
\end{tabular}

\label{tab2}
\end{table}

Table II shows the percentages of the different `seniority' (the number of broken quasiparticles) admixtures in the QRPA ground states obtained with the Bonn CD NN force for different nuclei and the basis with 9 and 13 single-particle levels defined in the caption of table I. The formalism for the determination of the relative percentages for the different seniorities  is described in ref.~\cite{Sanderson}. Making use of eqs.~(\ref{Q}),(\ref{BCS0}) and (\ref{X}), the admixtures of different `seniorities' (the number of broken quasiparticles) can be determined by the following eqs.~(\ref{C1}) to (\ref{s16}). The excitation of $ 0^+ $ unbroken  quasiparticle pairs should be excluded in the sums of eq. (\ref{s4}).

\begin{equation}
Y_{\alpha}^{m} = \sum_{\beta} C_{\alpha \beta} X_{\beta}^m, \ \ 
\label{C1}
C_{\alpha \beta} = \sum_m Y_{\alpha}^m (X_{\beta}^m)^{-1},
\end{equation}

\begin{equation}
w(s=4) = \frac{1}{4} \sum_{\alpha \beta} |C_{\alpha \beta}|^2
\label{s4}
\end{equation}

\begin{equation}
w(s=8) = \frac{1}{2} w(s=4)^2;\ 
%
w(s=12) = \frac{1}{6} w(s=4)^3;\ 
%
w(s=16) = \frac{1}{24} w(s=4)^4
\label{s16}
\end{equation}

\begin{figure}[htb]
\includegraphics[scale=0.65]{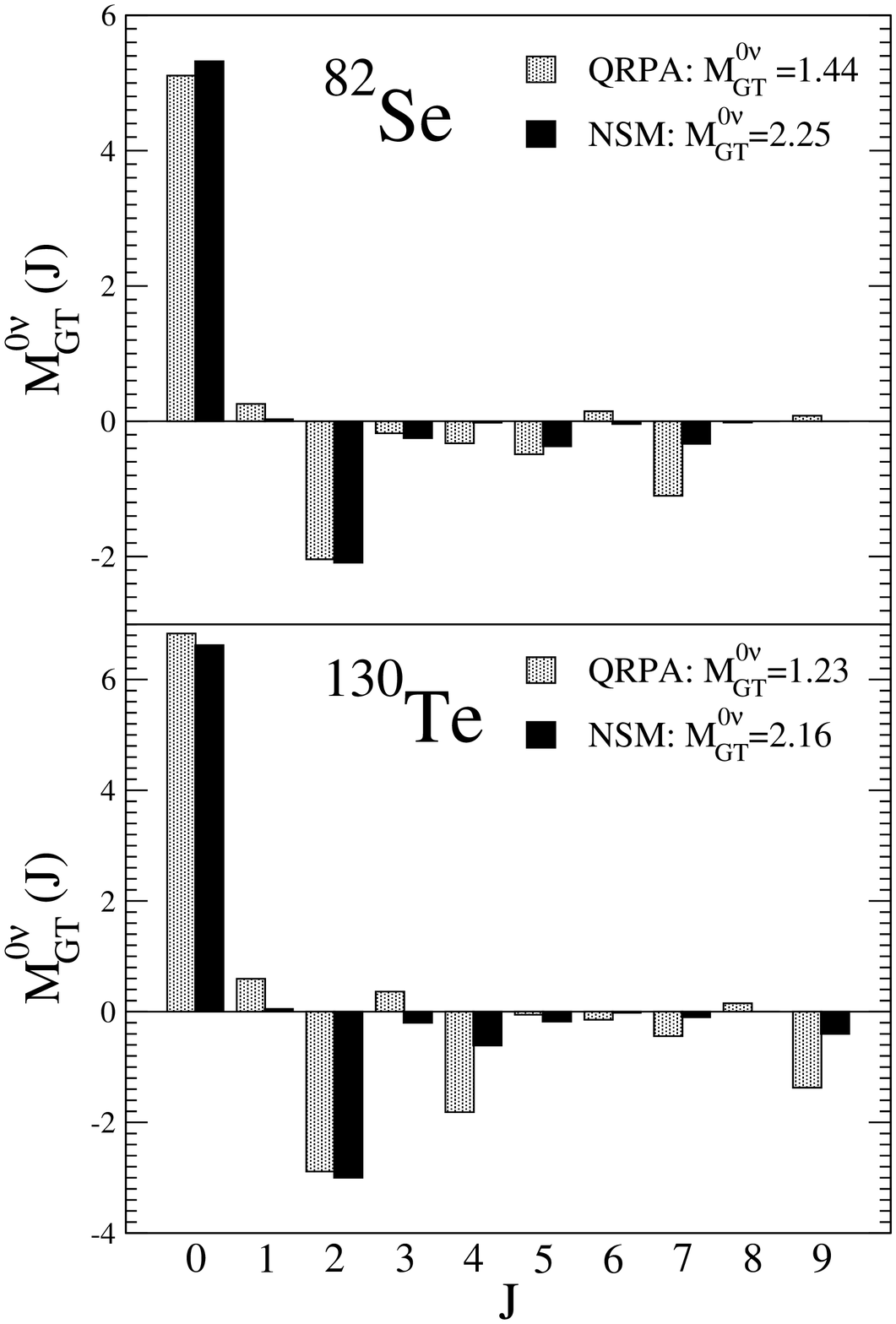}
\caption{Gamow Teller (involving the spin matrix $\sigma $) contributions of different angular momenta associated with the transforming neutron-pair for the QRPA and the SM in $^{82}$Se (upper panel) and  $^{130}$Te (lower panel). Since the Gamow Teller part is shown here only, the results are slightly different from figure ~\ref{QRPA-SM1} and table  ~\ref{tab1}, where the total matrix element is given.  The SM results shown in this figure do not include the effects of the higher order currents ~\cite{Poves1,Poves2}, while the QRPA ones do include them. The basis is the same small Shell Model basis for QRPA and the SM (four levels for $^{82}$Se and five for $^{130}$Te).}
\label{QRPA-SM}
\end{figure}

A comparison of the QRPA and the SM results for the Gamow Teller contributions of the different angular momentum pairs to the transition matrix elements is shown in figure ~\ref{QRPA-SM}. The basis has been chosen for the QRPA to be the same small basis as for the Shell Model. ( For $ ^{82}Se: 1p_{3/2},01f_{5/2}, 1p_{1/2}, 0g_{9/2} $ and for $ ^{130}Te:$ for protons $ 0g_{9/2}, 0g_{7/2},1d_{5/2},1d_{3/2}, 2s_{1/2}$ and for neutrons  $0g_{7/2}, 1d_{5/2}, 1d_{3/2}, 2s_{1/2} ,0h_{11/2}.)$ \\

In the last ten years P. K. Rath and coworkers ~\cite{Rath1, Rath2, Rath3} have published a whole series of papers (see references in Ref.~\cite{Rath2}) on $2\nu\beta\beta$ decay and, since 2008, also on $0\nu\beta\beta$ decay, in which they used a simple pairing plus quadrupole many body Hamiltonian of the Kumar and Baranger type~\cite{Kumar} to calculate the neutrinoless double beta decay transition matrix elements with angular momentum projection from a Hartree-Fock-Bogoliubov (HFB) wave function after variation. Schmid ~\cite{Schmid2} did show, that with the assumption of a real Bogoliubov transformation (real coefficient A and B)

\begin{equation}
a_{\alpha}^{\dagger} = \sum_{i = 1}^M A_{i \alpha}c_{i}^{\dagger}  + B_{i \alpha} c_i
\label{Bogo}
\end{equation}
and no parity mixing, only $0^+, 2^+, 4^+, \dots.$ nucleon pairs and excited states are allowed. In this approach one can describe rotational bands and can change neutron pairs with angular momentum $0^+, 2^+, 4^+, \dots$ into a corresponding  proton pair with the same angular momentum and parity in neutrinoless double beta decay for the ground to ground state transition in even mass nuclei. The proof uses the Bloch-Messiah ~\cite{Bloch} decomposition of the Bogoliubov transformation. After the first Bloch-Messiah transformation
\begin{equation}
b_{\alpha}^{\dagger} = \sum _{i}^M D_{i \alpha} c_{i}^{\dagger},
\label{Bloch1}
\end{equation}
one can write the HFB state as a BCS wave function:
\begin{equation}
|HFB\rangle = \prod_{\alpha, m_{\alpha} > 0}
[ u_{\alpha} + v_{\alpha} b_{\alpha}^{\dagger} b_{\bar{\alpha}}^{\dagger}] |0\rangle
\label{BCS}
\end{equation}

The bar indicates the time reversed nucleon state. The neutron pairs to be changed into two protons in a $0^+ \to 0^+$ $0\nu\beta\beta$ transition can be written for axial-symmetric deformed nuclei:
\begin{eqnarray}
b_{\alpha}^{\dagger} b_{\bar{\bar{\alpha}}}^{\dagger} = \sum_{\tau = p,n}\sum_{i \le k} [1 + \delta_{i,k}]^{-1} \sum_{J} (-)^{l_k + j_k -  m_{\alpha}} (j_i,j_k, J|m_{\alpha}, -m{\alpha},0) \nonumber\\
\cdot [ \Re (D_{i, \alpha}^* D_{k, \alpha}) [ 1 +  (-)^{l_i +l_k +J} ]
+ i \Im (D_{i, \alpha}^* D_{k, \alpha}) [ 1 -  (-)^{l_i +l_k +J} ] ][c_i^{\dagger} c_k^{\dagger}]^{J, 0}_{1, 2\tau} \\
\sum_{i,k} \sum_{J, T} (1/2, 1/2, T|-1/2, 1/2, 0)(-)^{l_k + j_k - m_{\alpha}} (j_i,j_k, J|m_{\alpha}, -m_{\alpha},0) \nonumber\\
\cdot [ \Re(D_{i \alpha}^* D_{k \alpha}) [ 1 +  (-)^{l_i +l_k +J} ]
+ i \Im (D_{i \alpha}^* D_{k \alpha}) [ 1 -  (-)^{l_i +l_k +J} ] ][c_i^{\dagger} c_k^{\dagger}]^{J, 0}_{T,0}.\nonumber
\label{pair1}
\end{eqnarray}

If only real coefficients are admitted and no parity mixing is allowed in the Bogoliubov transformation (\ref{Bogo})~\cite{Rath1,Tomoda1,Rath2}, the expressions for the neutron pairs to be transformed into a proton pair for a $0^+ \to 0^+$ transition reduces to:
\begin{eqnarray}
b_{\alpha}^{\dagger} b_{\bar{\alpha}}^{\dagger} = \sum_{i \le k} [1 + \delta_{i,k}]^{-1} \sum_{J} (-)^{l_k + j_k - m_{\alpha}} (j_i,j_k, J|m_{\alpha}, -m{\alpha},0)
\cdot [ (D_{i \alpha}^* D_{k \alpha}) [ 1 +  (-)^{J} ]]
[c_i^{\dagger} c_k^{\dagger}]^{J, 0}_{1, 2\tau}.
\label{pair2}
\end{eqnarray}

The neutron pairs, which can be transformed into a proton pair for $0^+ \to 0^+$  transitions are restricted to angular momenta $0^+, 2^+, 4^+, 6^+, \dots $.

\begin{figure}[htb]
\includegraphics[scale=0.65]{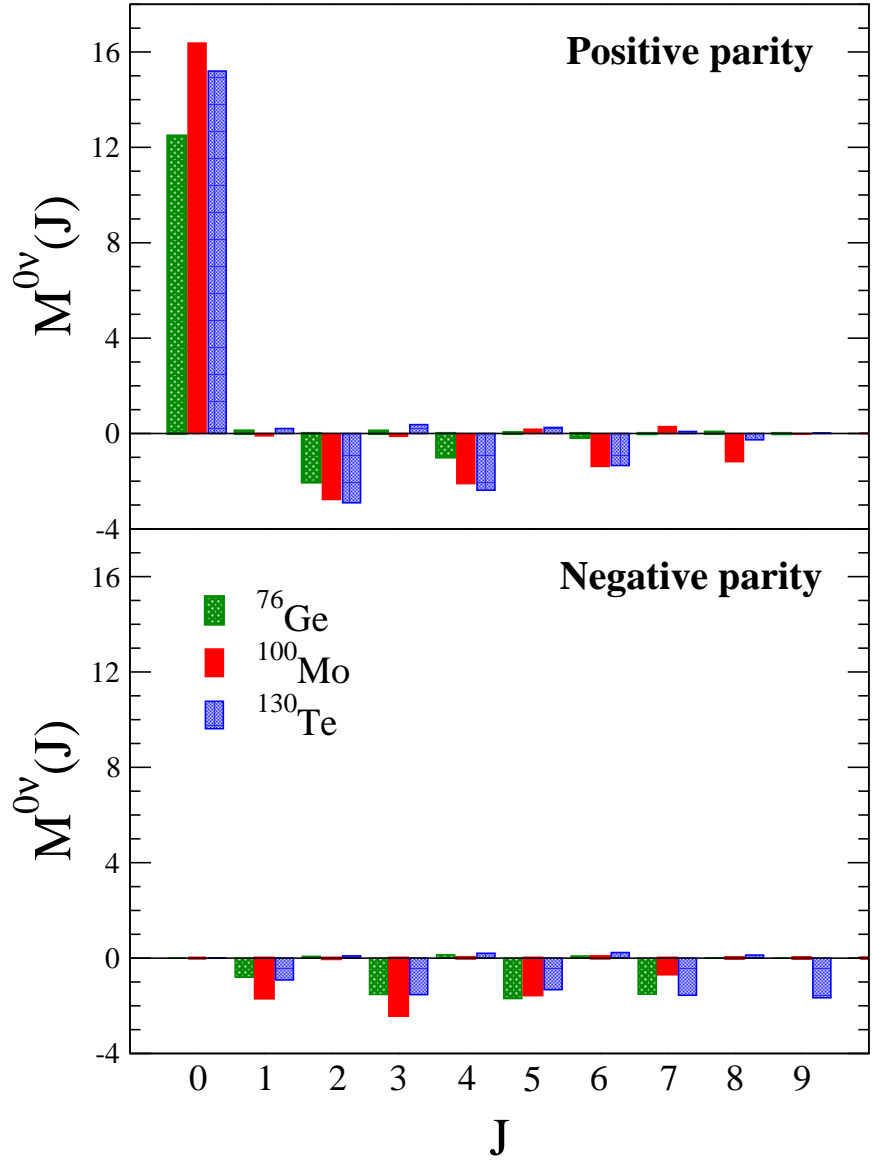}
\caption{(Color online)  Contribution of different angular momenta for both parities (positive above and negative below)  of the decaying neutron pairs in  QRPA for $^{76}$Ge, $^{100}$Mo and $ ^{130}$Te. The radius parameter $r_0$ is chosen to be 1.2 fm as in the whole publication.}
\label{Ge-Mo-Te}
\end{figure}

The QRPA and the SM do not have this restriction. Figure ~\ref{Ge-Mo-Te} shows  the contributions of the different angular momentum and parity pairs to the $0\nu\beta\beta$ matrix elements in $^{76}$Ge, $^{100}$Mo and $^{130}$Te calculated within the QRPA.

\begin{figure}[htb]
\includegraphics[scale=0.65
]{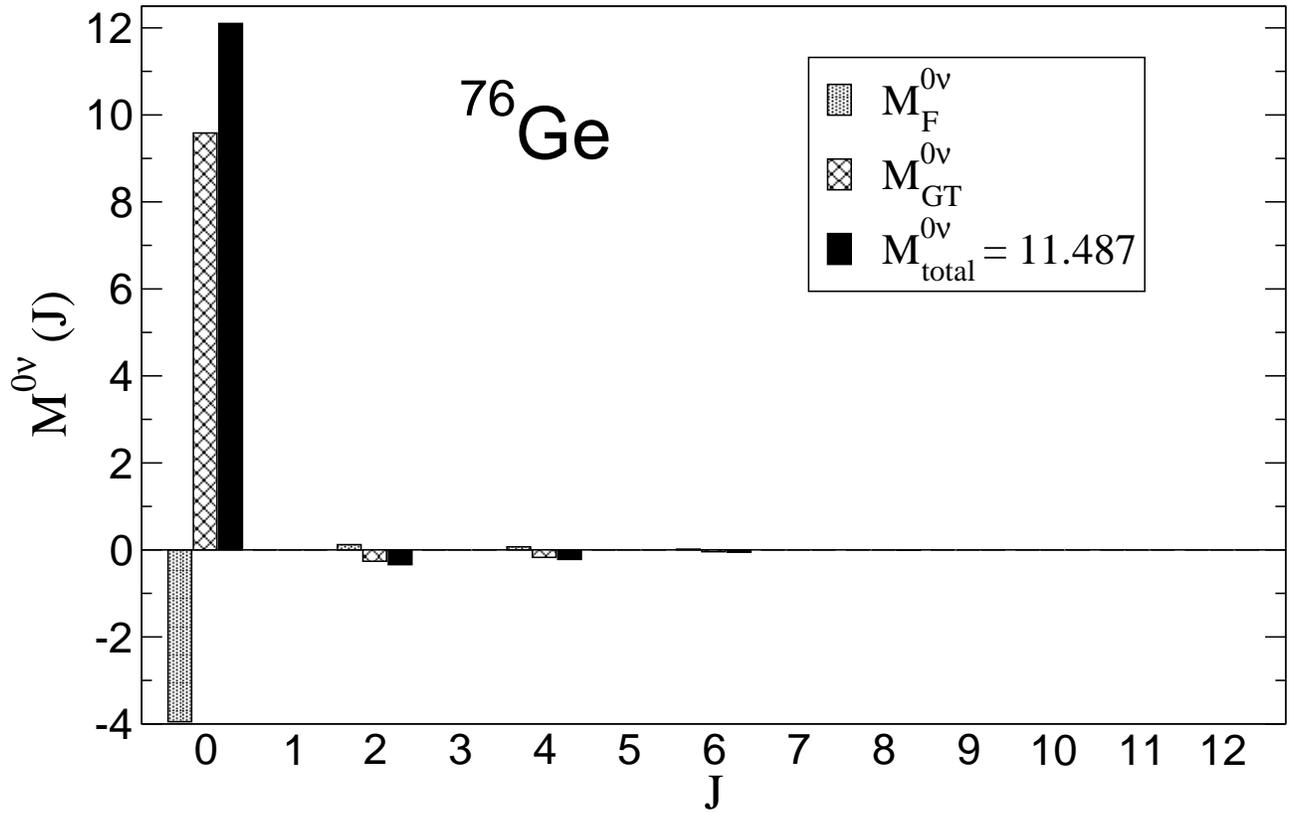}~~
\caption{
Contributions of neutron pairs with different angular momenta to the neutrinoless double beta decay transition matrix elements for $^{76}$Ge $\to ^{76}$Se calculated from a HFB wave function with angular momentum and proton and neutron particle number projection before variation. The Fermi, the Gamow-Teller and the total contribution including the tensor part as defined in eq.~(\ref{Mnu}) are separately given. The nucleon-nucleon interaction is an improved Gogny type force~\cite{Gogny}. The deformations $\beta_{Ge} =-0.08$ and $\beta_{Se} = 0.11$ correspond to the minima of the projected HFB total energy. The results are qualitatively and almost quantitatively the same for the experimental deformation from the Coulomb reorientation effect: $ \beta_{Ge} = 0.16, \beta_{Se} = 0.10 $ and also for different forces. The angular momenta of the  neutron pairs are restricted  to $0^+, 2^+, 4^+, \dots $. In addition the contributions of higher angular momentum neutron pairs $2^+, 4^+, \dots $ are drastically reduced compared to the QRPA and the SM. }
\label{HFB-Fig}
\end{figure}

Figure \ref{HFB-Fig} shows on the other side, that the projected HFB approach is restricted to contributions of neutron pairs with angular momenta $0^+, 2^+, 4^+, \dots $. In addition, one sees that the contributions of transition of higher angular momentum neutron to proton pairs $2^+, 4^+, \dots$  are drastically reduced compared to the QRPA and the SM see fig. \ref{HFB-Fig}. The reason for this is obvious: in a spherical nucleus the HFB solution contains only seniority zero and no stronger higher angular momentum pairs. The double beta decay system considered has only a small deformation and thus a projected HFB state is not able to describe an appreciable admixture of higher angular momentum pairs for $0^+ \to 0^+$ transitions as can be seen from the wave function (\ref{BCS}) after the first Bloch-Messiah transformation \cite{Bloch}. The higher angular momentum contributions increase drastically with increasing intrinsic quadrupole and hexadecapole deformations of the HFB solution. \\

The results in figure ~\ref{HFB-Fig} are calculated by K.W. Schmid~\cite{Schmid3} within the HFB with the angular momentum projection before variation with an improved Gogny force~\cite{Gogny} adjusted in a global fit to properties of many nuclei.

To have also $1^+, 3^+, 5^+, \dots $ neutron pairs contributing one has to use a Bogoliubov transformation  with complex coefficients A and B [see equations (\ref{Bogo}) and (\ref{pair1})]. To have also $0^-, 1^-, 2^-, 3-, 4^-, 5^-, \dots $ one has to allow parity mixing in the Bogoliubov transformation ( \ref{Bogo}). But even allowing all different types of angular momentum and parity pairs one would still have an unnatural suppression of the higher angular momenta especially for smaller deformations. This handicap could probably be overcome by a multi-configuration HFB wave function \cite{Schmid2} with complex coefficients and parity mixing in the Bogoliubov (\ref{Bogo}) transformation.

The IBM (Interacting Boson Model)~\cite{Iachello} can only change $0^+$ (S) and $2^+ $ (D) fermionic pairs from two neutrons into two protons. In the bosonization to higher orders this leads to the creation and annihilation of up to three ``s'' and ``d'' boson annihilation and creation operators in Ref.~\cite{Iachello}. But all these terms  of equation (18) of reference~\cite{Iachello} originate from the annihilation of a $0^+$ (S) or a $2^+ $ (D) neutron pair into a corresponding proton pair with the same angular momentum. The higher boson terms try only to fulfill the Fermi commutation relations of the original nucleon pairs up to third order. The IBM can therefore change only a $0^+$ or a $2^+$ neutron pair into a corresponding proton pair.

\section{Conclusions}

The Shell Model (SM) \cite{Poves1, Poves2, Poves3} is in principle the best method to calculate the nuclear matrix elements for neutrinoless double beta decay. But due to the restricted single-particle basis it has a severe handicap. The matrix elements in the $^{76}Ge$ region are by a factor 2  smaller than the results of the Quasiparticle Random Phase Approximation (QRPA)~\cite{Rod05, Ana, Suh05}. With the same restricted basis as used by the SM the QRPA obtains roughly the same results as the SM (see figure 5), but the Ikeda sum rule~\cite{Ikeda} gets strongly violated due to the  spin-orbit partners missing in the SM single-particle basis. Future increases in computer performance and perhaps also more powerful methods of programming might bring improvements for the SM calculations of $M^{0\nu}$.

The SM can admix seniority $ 0, 4, 6, 8, 10, \dots$ configurations to the ground state, while the QRPA is restricted in the ground state to the "seniority" (the number of broken quasiparticles)  $ 0, 4, 8, 12, \dots $ configurations (see figures ~\ref{ring} and  \ref{Poves-Fig}). But the SM results show also, that the seniority 6 and 10 configurations, which are not included in the QRPA ground state,
contribute just a little to the total
neutrinoless double beta decay matrix element (see figure ~\ref{Poves-Fig}).

The angular momentum projected Hartee-Fock-Bogoliubov (HFB) method \cite{Rath1} is restricted in its scope. With a real Bogoliubov transformation without parity mixing (\ref{Bogo}) one can only describe neutron pairs with angular momenta and parity $ 0^+, 2^+, 4^+, 6^+, \dots $ changing into two protons for ground state-to-ground state transitions. The restriction for the Interacting Boson Model (IBM) \cite{Iachello} is even more severe: one is restricted to $0^+$ and $2^+$ neutron pairs changing into two protons.

\begin{figure}[t]
\includegraphics[scale=0.6
]{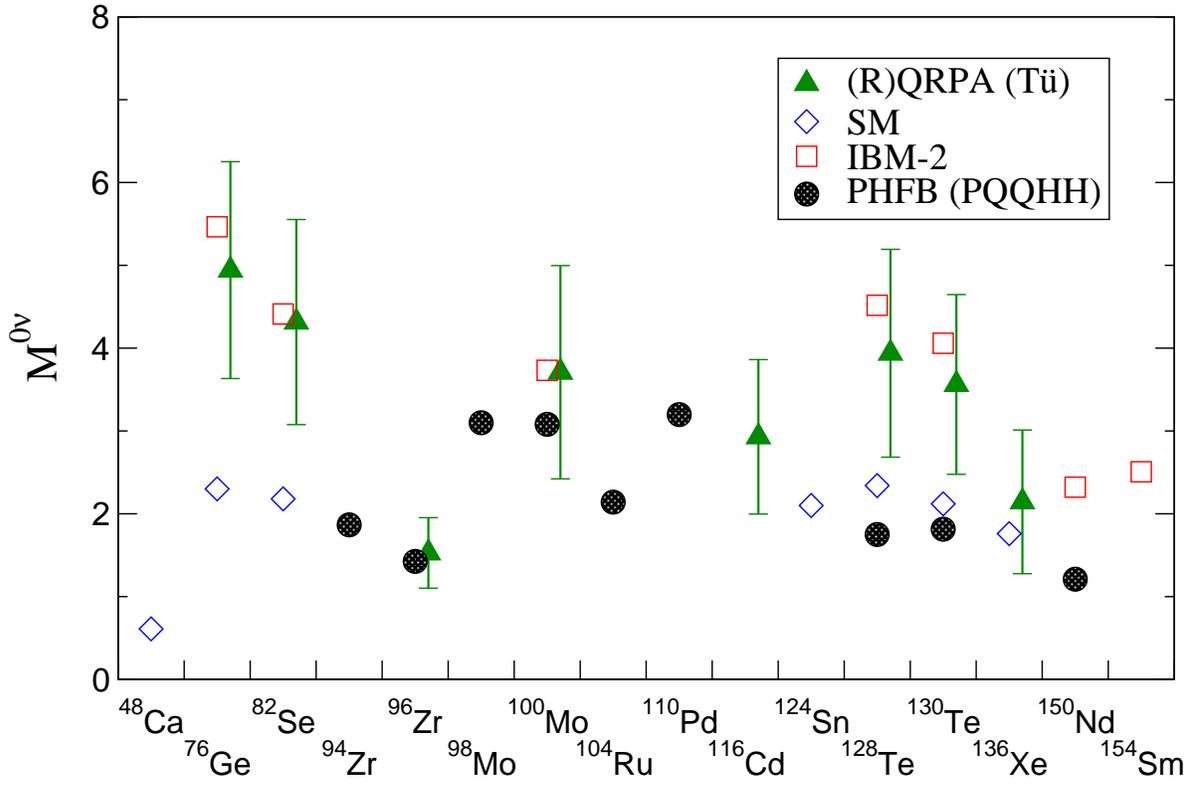}
\caption{(Color online) \ Neutrinoless double beta decay transition matrix elements for the different approaches: QRPA~\cite{Rod05, Ana}, the SM 
\cite{Poves1, Poves2, Poves3}, the projected HFB 
method~\cite{Rath3} and the IBM
\cite{Iachello}. The error bars  for the QRPA are calculated as the highest and the lowest values for three different single-particle basis sets, 
two different axial charges $g_A = 1.25$ and the quenched value $g_A = 1.00$ and two different treatments of short range correlations (Jastrow-like~\cite{Miller} and the Unitary Correlator  Operator Method (UCOM)~\cite{Feld}). The radius parameter is as in this whole work $r_{0} = 1.2$ fm. }
\label{total}
\end{figure}

A comparison of the $0\nu\beta\beta$ transition matrix elements calculated recently in the different many body methods: QRPA, SM, HFB and IBM is shown in Fig.~\ref{total}.

\acknowledgments
The authors acknowledge support of the Deutsche Forschungsgemeinschaft within both the SFB TR27 "Neutrinos and Beyond" and the Graduiertenkolleg GRK683.

\newpage

\end{document}